\documentclass[aps,prx,reprint,superscriptaddress]{revtex4-1}

\usepackage{graphicx,hyperref,amsmath,epstopdf,natbib,textcomp, multirow, array}

\begin{document}

%\preprint{}

\title{Large Stark tuning of donor electron spin quantum bits in germanium}



\author{A. J. Sigillito}
\email[]{asigilli@princeton.edu}
%\thanks{}
\affiliation{Department of Electrical Engineering, Princeton University, Princeton, New Jersey 08544, USA}

\author{A. M. Tyryshkin}
%\thanks{}
\affiliation{Department of Electrical Engineering, Princeton University, Princeton, New Jersey 08544, USA}

\author{J. W. Beeman}
%\thanks{}
\affiliation{Materials Sciences Division, Lawrence Berkeley National Laboratory, Berkeley, California 94720, USA}

\author{E. E. Haller}
%\thanks{}
\affiliation{Department of Materials Science and Engineering, University of California, Berkeley, California 94720, USA}
\affiliation{Materials Sciences Division, Lawrence Berkeley National Laboratory, Berkeley, California 94720, USA}

\author{K. M. Itoh}
%\thanks{}
\affiliation{School of Fundamental Science and Technology, Keio University, 3-14-1 Hiyoshi, Kohuku-ku, Yokohama 223-8522, Japan}

\author{S. A. Lyon}
%\thanks{}
\affiliation{Department of Electrical Engineering, Princeton University, Princeton, New Jersey 08544, USA}


\date{\today}

\begin{abstract}

Donor electron spins in semiconductors make exceptional quantum bits because of their long coherence times and compatibility with industrial fabrication techniques. Despite many advances in donor-based qubit technology, it remains difficult to selectively manipulate single donor electron spins. Here, we show that by replacing the prevailing semiconductor host material (silicon) with germanium, donor electron spin qubits can be electrically tuned by more than an ensemble linewidth, making them compatible with gate addressable quantum computing architectures. Using X-band pulsed electron spin resonance, we measured the Stark effect for donor electron spins in germanium. We resolved both spin-orbit and hyperfine Stark shifts and found that at 0.4 T, the spin-orbit Stark shift dominates. The spin-orbit Stark shift is highly anisotropic, depending on the electric field orientation relative to the crystal axes and external magnetic field. When the Stark shift is maximized, the spin-orbit Stark parameter is four orders of magnitude larger than in silicon. At select orientations a hyperfine Stark effect was also resolved and is an order of magnitude larger than in silicon. We report the Stark parameters for $^{75}$As and $^{31}$P donor electrons and compare them to the available theory. Our data reveal that $^{31}$P donors in germanium can be tuned by at least four times the ensemble linewidth making germanium an appealing new host material for spin qubits that offers major advantages over silicon.

\end{abstract}

%\keywords{}

\maketitle

\section{Introduction}

Since the 70's, silicon-based computing has roughly followed Moore's law, doubling the density of transistors on a chip (and effectively the computing power) approximately every two years. As the transistor size approaches the atomic limit\cite{fuechsle2012}, research has turned from miniaturizing transistors towards replacing silicon with higher performance materials like germanium\cite{shang2003, yu2011, kim2014, scappucci2011}. In parallel, the field of quantum information processing has been innovating the way computers work by taking advantage of quantum effects. Bolstered by the semiconductor industry, the field of donor spin qubits in semiconductors has rapidly advanced over the past decade and now even single-donor devices can be fabricated\cite{Pla2012, Pla2013}. Donor-based spin qubits are not only compatible with industrial fabrication techniques, they also boast long coherence times\cite{tyryshkin2012, SigillitoGe2015} and are easy to control. Even so, the ability to reliably perform local, single-qubit operations in a scalable architecture--a prerequisite for universal quantum computing\cite{Divincenzocriteria}--has remained elusive. Spin qubits are typically manipulated using resonant microwave magnetic field pulses, but the fields are difficult to confine at the single-spin scale. The conventionally proposed solution is to use local electrical gates to tune individual spins on and off resonance with a globally applied microwave magnetic field\cite{kane1998, hill2005}. This electric field induced shift in the spin resonance frequency is known as the Stark effect. This effect has been measured for donors in silicon\cite{bradbury2006, lo2014, pica2014, sigillitoStark2015} and is weak; it is unable to shift the resonance more than a small fraction of the inhomogeneous linewidth. Here, we overcome this problem by substituting the ubiquitous semiconductor, silicon, with germanium. Germanium is a fundamentally different material that supports long coherence times \cite{SigillitoGe2015} and is gaining popularity for spintronics applications\cite{li2013, dushenko2015, shen2010, giorgioni2016}. We measured the Stark effect for donors in germanium and found that it is substantially larger than in silicon. The Stark effect for donors in germanium is comprised of two parts: the hyperfine and spin-orbit Stark shifts which are respectively one and four orders of magnitude larger than in silicon. This means that for even the small electric fields applied in this work (480 V/cm), $^{31}$P donor electron spins in Ge can be tuned by at least four times the ensemble linewidth at X-band magnetic fields. This pioneering work shows that germanium is a promising new host material for the next generation of donor spin qubits.

\begin{figure} [h]
\includegraphics{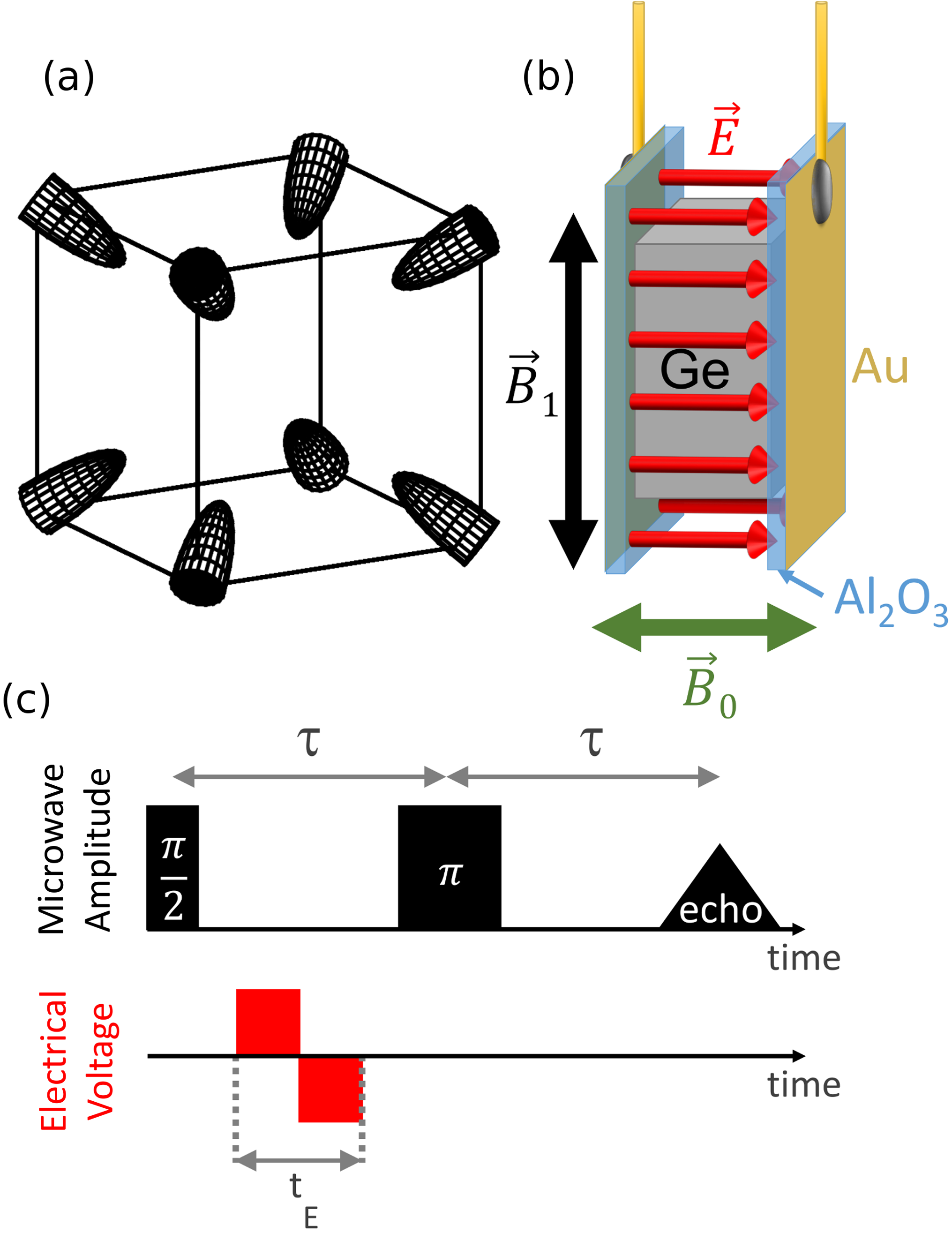}
\caption{\label{fig:fig1exp} (a) Cartoon illustrating the valley structure in germanium (half-ellipsoids) superimposed on a unit cell of the crystal. The sides of the cube are oriented along the (100) equivalent crystallographic directions. (b) Cartoon of the parallel plate capacitor scheme used for applying electric fields to the Ge samples. The electric field (red) is uniform over the sample volume and directed between the two Au electrodes. When placed in the microwave resonator, $\vec{B}_{1}$ (black) is directed up and down and $\vec{B}_{0}$ (green) can be oriented in any arbitrary direction orthogonal to $\vec{B}_{1}$. (c) Schematic representation of the pulse sequence used to measure Stark shifts. Microwave pulses are shown in black whereas the bipolar electric field pulse is shown in red.}
\end{figure}

While this is the first experimental work looking at the Stark tunability of donors in germanium, there have been several measurements made in silicon\cite{bradbury2006, lo2014, pica2014, sigillitoStark2015}. Most studies have directly measured the Stark effect for donors in silicon, but a few experiments have demonstrated the Stark addressability of donor qubits under certain conditions. Stark addressability has been shown for narrow linewidth nuclear spin ensembles \cite{wolfowicz2014nuc}, but the shifts are too small to address electron spin ensembles. Stark tuning an individual donor electron spin on and off resonance with a driving microwave field has also been demonstrated \cite{lauchte2015} but is insufficient for multi donor quantum computing schemes where each donor will experience a different inhomogeneous environment. Based on the direct measurements of the Stark parameters \cite{bradbury2006, lo2014, pica2014, sigillitoStark2015}, it remains unlikely that one can tune donor electron spins by more than an ensemble linewidth before ionization sets in.

Germanium's large tunability arises from its large spin-orbit coupling, small valley-orbit splitting, and small binding energy \cite{wilson1964}. Our measurements of the Stark effect for donor electron spins ($^{31}$P and $^{75}$As) in germanium can be well understood by the effective mass theory of Pica \textit{et al.}\cite{picainprep} and the multimillion atom tight binding theory of Rahman \textit{et al.} \cite{rahman2009} which are in good agreement with our data. Physically, the hyperfine Stark effect arises from a shift in the electronic wave function away from the donor nucleus when an electric field is applied. The hyperfine coupling is proportional to the overlap of the electronic wave function with the donor nucleus, so a shift in the wave function results in a reduction in the hyperfine coupling. This is the smaller of the two effects at X-band magnetic fields. The second contribution to the overall Stark shift is the spin-orbit Stark effect, which arises from a modulation of the electron wavefunction in the conduction band valleys, thus affecting the g-factor \cite{wilson1964, feher1959}.

The conduction band valley structure controls the spin-orbit Stark effect in germanium. Germanium has 4 valley ellipsoids (or 8 half ellipsoids) centered at the L-points of the Brillouin zone (along the $<$111$>$ equivalent crystallographic axes)\cite{roth1960, hasegawa1960}. This is depicted by the cartoon in Fig.~\ref{fig:fig1exp}(a). Each individual valley has a highly anisotropic $g$-factor with values varying from 1.92 to 0.82 for $^{75}$As donors (or 1.93 to 0.83 for $^{31}$P donors)\cite{wilson1964}. The donor ground state is a weighted superposition of the four valleys, and therefore the overall $g$-tensor is given by a weighted sum over all of the individual valley $g$ tensors. This gives
\begin{equation} \tensor{g}_{eff} = \sum\limits_{i=1}^{4}{\alpha_{i}\tensor{g_i}}
\end{equation}
where  $\tensor{g}_{eff}$ is the overall $g$-tensor, $\alpha_{i}$ is the wavefunction amplitude in the $i$-th valley, and $\tensor{g_i}$ is the $g$-tensor of an individual valley\cite{feher1959}. Each valley has an axially symmetric $g$-tensor given as
\begin{equation}
\tensor{g_{i}} = \begin{bmatrix}  g_{\perp} & 0 & 0 \\ 0 & g_{\perp} & 0 \\ 0 & 0 & g_{\parallel} \end{bmatrix}
\end{equation}in the valley basis, with $g_{\perp}$ and  $g_{\parallel}$ equal to the $g$ factors perpendicular and parallel to the valley axis, respectively. In the absence of any electric fields or strain, the electron wave function equally populates the valleys ($\alpha_{i} = 0.25$) leading to an isotropic g-value ($\tensor{g_{eff}} = g_{0}I$ where $g_{0}$ = 1.57 for $^{75}$As, 1.5631 for $^{31}$P,  and $I$ is the identity matrix)\cite{wilson1964}. When an electric field is applied, the valleys with axes oriented along the electric field are lowered in energy, and their $\alpha_{i}$ increase relative to the other valleys. This gives rise to anisotropy in $\tensor{g_{eff}}$. In addition to this valley-repopulation effect, a g-factor shift can result from the ''single-valley" effect where an electric field mixes the ground state with higher lying conduction bands\cite{feher1959}. This can be thought of as a modulation of $g_{\parallel}$ and $g_{\perp}$ as opposed to the modulation of $\alpha_{i}$ caused by the valley repopulation effect.

From symmetry considerations, the Stark effect for donor electron spins must be quadratic to first order \cite{pica2014, rahman2009} so that the Stark-induced frequency shift, $df$, can be described as

\begin{equation}
df =  [\eta_{g} g \beta B_{0} + \eta_{A} A M_{I}] \vec{E}^{2}
\end{equation}
where $\eta_{g}$ and $\eta_{A}$ are the spin-orbit and hyperfine Stark parameters, respectively, $g$ is the g-factor along $\vec{B}_{0}$, $\beta$ is the Bohr magneton, $\vec{B}_{0}$ is the magnetic field, $A$ is the hyperfine coupling constant, $M_{I}$ is the nuclear spin projection, and $\vec{E}$ is the applied electric field. The spin-orbit Stark parameter includes both the valley repopulation and single valley Stark shifts and thus depends on the direction of the applied electric and magnetic fields. In this work, we measure the angular dependence of the Stark parameters for $^{75}$As and $^{31}$P donors and find that in certain orientations they are four orders of magnitude larger than what was measured for donors in silicon. These large Stark parameters indicate that germanium-based spin qubits have some important advantages over their silicon analogues. 

\section{Experimental Methods}

The Stark shift was measured using a pulsed electron spin resonance (ESR) technique sensitive to small frequency shifts, as described by Mims \cite{mims1974}. This technique uses a Hahn echo pulse sequence with an electric field pulse of length $t_{E}$ inserted between the microwave pulses as illustrated in Fig.~\ref{fig:fig1exp}(c). The applied electric field detunes the spins relative to the local oscillator of the microwave bridge such that they accumulate a phase, $d\phi$, which is readily measured using a quadrature detector. The phase shift is directly related to the Stark shift ($df$) by $df = d\phi/ t_{E}$. To cancel linear Stark effects, which can arise from strain\cite{bradbury2006, pica2014, sigillitoStark2015}, bipolar electric field pulses were used as described in Ref.~\cite{bradbury2006}.

Five samples were measured in this work, and their details are outlined in Table~\ref{table:samples}. Three of the samples are commercially available natural germanium, and two other crystals are isotopically enriched \cite{itoh1993, itoh1994}. The isotopic enrichment is particularly important for these experiments because it allows the donor hyperfine structure of the ESR spectra to be resolved. In natural germanium, hyperfine interactions with the spin 9/2 $^{73}$Ge nuclei (7.8\% abundant) broaden the lines to the extent that the donor hyperfine structure and thus the hyperfine Stark shifts cannot be clearly resolved. The natural Ge samples were therefore only used to measure the spin-orbit Stark parameters. The first isotopically enriched crystal is primarily $^{74}$Ge and contains approximately 3.8\% $^{73}$Ge \cite{itoh1993, itoh1994, SigillitoGe2015}. This crystal was neutron transmutation doped to a density of $3 \times  10 ^{15}$ $^{75}$As donors/cm$^{3}$. The other isotopically enriched sample is a piece of $^{70}$Ge that only contains 0.1\% $^{73}$Ge and has approximately $10^{12}$ $^{31}$P donors /cm$^{3}$.

\begin{table}[h]
\caption{Sample Details} % title of Table
\centering % used for centering table
\begin{tabular}{c c c c c c}
\hline\hline %inserts double horizontal lines
 Number & Material  & Doping  &  Faces & & $T_{2}$   \\ [0.5ex]
%heading
\hline % inserts single horizontal line
1&$^{74}$Ge:As & $3 \times 10^{15}$ As/cm$^{3}$ & $[110]$ & $[001]$ & 114 $\mu$s \\ % inserting body of the table
2&$^{nat}$Ge:As & $1 \times 10^{15}$ As/cm$^{3}$ & $[\overline{1}11]$ & $[01\overline{1}]$ & 55 $\mu$s  \\
3&$^{70}$Ge:P &  $\sim 10^{12}$ P/cm$^{3}$ & $[100]$ & $[001]$ & 250 $\mu$s \\
4&$^{nat}$Ge:P &  $4 \times 10^{14}$ P/cm$^{3}$ & $[110]$ & $[001]$ & 55 $\mu$s \\
5&$^{nat}$Ge:P &  $10^{13}$ P/cm$^{3}$ & $[111]$ & $[1\overline{1}0]$ & 55 $\mu$s  \\
\hline %inserts single line
\end{tabular}
\label{table:samples} % is used to refer this table in the text
\end{table}

All of the samples were cut to have faces along primary crystal axes (outlined in Table~\ref{table:samples}) and X-ray diffraction was used to verify that all faces were within approximately 1$^{\circ}$ of the intended planes. These faces were used to align the electric field to the crystal. Additionally, the magnetic field must be aligned to the crystal so the sample holder was equipped with a goniometer. The goniometer was calibrated to within $\sim2^{\circ}$ by measuring the ESR linewidth as a function of angle since the linewidth is minimized for $B_{0}$ in the (100) direction \cite{wilson1964}.

To apply uniform electric fields, samples were sandwiched between gold electrodes in a parallel plate capacitor arrangement as shown in Fig.\ref{fig:fig1exp}(b). The electrodes were fashioned from double side polished sapphire wafers with 200 nm of gold deposited on the surface. It was necessary to keep the gold layers thin to avoid loading the microwave resonator. The samples were secured in the parallel plate structures by loosely wrapping them in teflon tape before inserting them into an X-band dielectric resonator (Bruker MD-5) equipped with a low noise cryogenic preamplifier. The samples were cooled to 1.8 K in a pumped helium cryostat.

We measured the Stark shift at 9.6 GHz using 200 ns and 400 ns $\pi/2$ and $\pi$ pulses and a resonator Q factor of 2000. All experiments were conducted at 1.8 K where the samples have conveniently short spin-lattice relaxation times, $T_{1}$ $\sim$1 ms \cite{SigillitoGe2015}. The spin echoes were typically signal averaged 1000 times per experiment and every experiment was repeated 50 times to further improve the signal to noise ratio. The dephasing time, $\tau$, was kept short relative to the coherence time, $T_{2}$, as given in Table~\ref{table:samples} \cite{SigillitoGe2015}. This sets a limit on the length of the electric field pulse, $t_{E}$. For the $^{nat}$Ge samples $t_{E}$ was typically 10 $\mu$s while $t_{E}$ of 30-45 $\mu$s were used for the isotopically enriched samples.

\section{Results and Discussion}

\begin{figure} [h]
\includegraphics{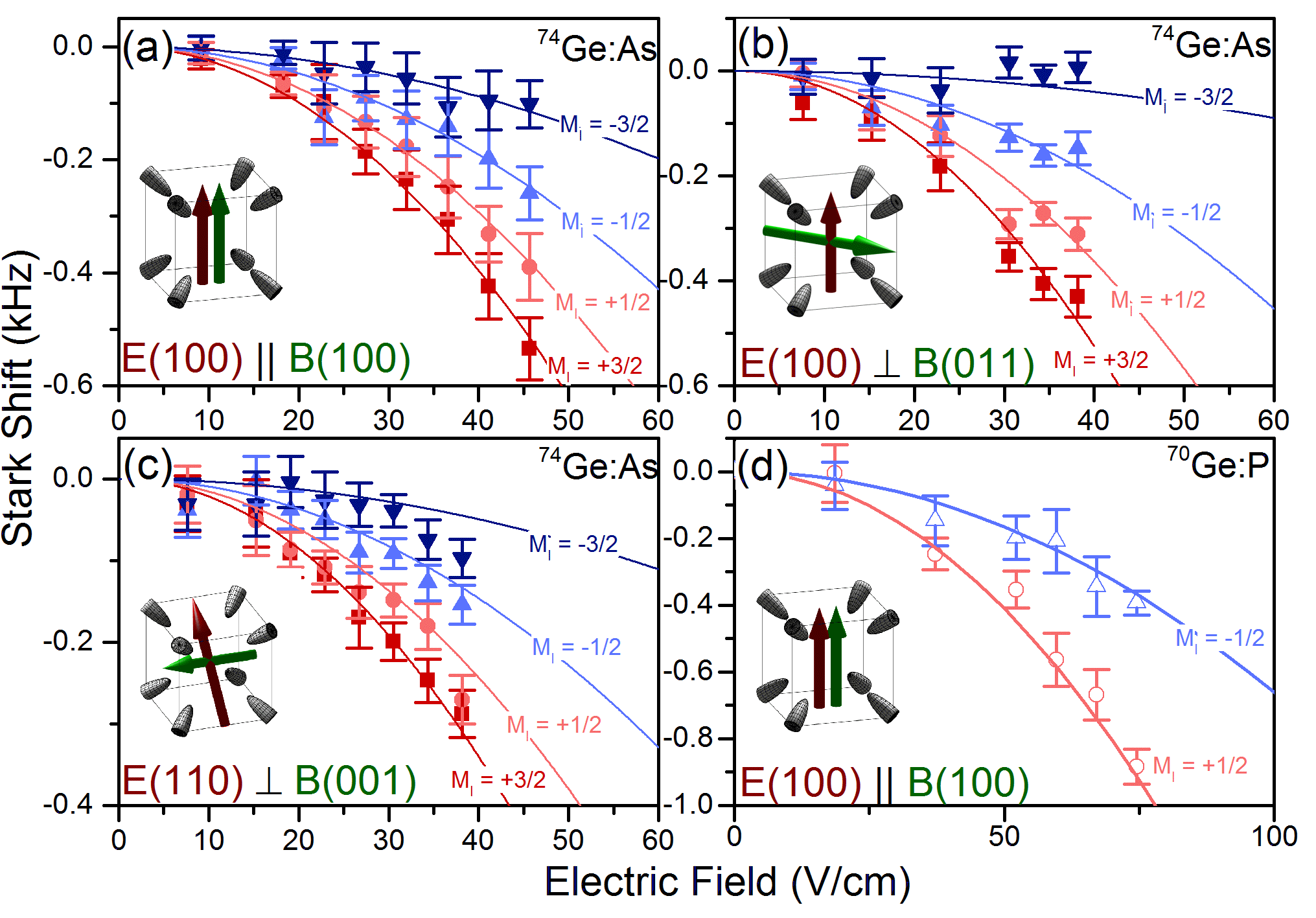}
\caption{\label{fig:fig2exp} Stark shifts measured for $^{75}$As (a-c) and $^{31}$P (d) (samples 1 and 3, respectively) for different configurations of the electric and magnetic fields (denoted by the cartoon insets). The red arrow indicates the direction of the electric field whereas the green arrow shows the direction of $B_{0}$ relative to the conduction band valleys (grey ellipsoids). In the plots, different symbols/colors denote different hyperfine lines ($M_{I}$). The fanning out of the Stark shifts comes from the hyperfine Stark effect whereas the center of mass shift comes from the spin-orbit Stark effect. Solid lines represent the global least squares fit to the data using Eq.(3) with fitting parameters listed in Table 2.}
\end{figure}

The Stark shift data for the $^{74}$Ge:As and $^{70}$Ge:P samples are plotted in Fig.~\ref{fig:fig2exp} for various electric and magnetic field configurations. These were the only samples where we were able to clearly resolve all four $^{75}$As or two $^{31}$P donor hyperfine lines ($M_{I}$) and the measurements were performed on each line. The data clearly resolve both hyperfine (fanning out) and spin-orbit (center of mass shift) Stark effects. In Figs.~\ref{fig:fig2exp}(a-b) and (d) the electric field is oriented along a (100) crystallographic direction that makes equal angles with all four conduction band valleys (oriented in the (111) equivalent directions). In this orientation, there should be no valley repopulation because all valleys experience the same energy shift. This means that only the single-valley effect is responsible for the observed Stark shifts. For Fig.~\ref{fig:fig2exp}(c), the electric field makes different angles with the valleys, and therefore both valley repopulation and single-valley Stark effects can occur. However, here the magnetic field makes an equal angle with all of the valleys. In this configuration, each valley has an equivalent $g$ factor and redistribution of the electronic wave function among the valleys cannot affect $\tensor{g}_{eff}$. The data for $\vec{E} \ || \ \vec{B}_{0} \ || \ (110)$ has been omitted from Fig.~\ref{fig:fig2exp} since the signal-to-noise was too poor to resolve the hyperfine component of the Stark shift. We fit Eq.~3 to the data and report the extracted Stark parameters in Table 2. Since neither of the isotopically enriched crystals have faces cut in the (111) direction, we could not measure the Stark effect for E$\parallel$(111) in these crystals.

Natural germanium crystals were available with faces cut in all of the primary crystals planes, but since they have broad ESR linewidths, we were not able to measure the hyperfine Stark shift and only measured the spin-orbit Stark shift. To accurately determine the spin-orbit term, the Stark shift was measured at the expected center of each hyperfine line and the results were averaged. Because the hyperfine Stark shift is proportional to $M_{I}$, averaging over opposite hyperfine lines cancels out the hyperfine Stark shift so that only the spin-orbit term survives.

\begin{figure} [h]
\includegraphics{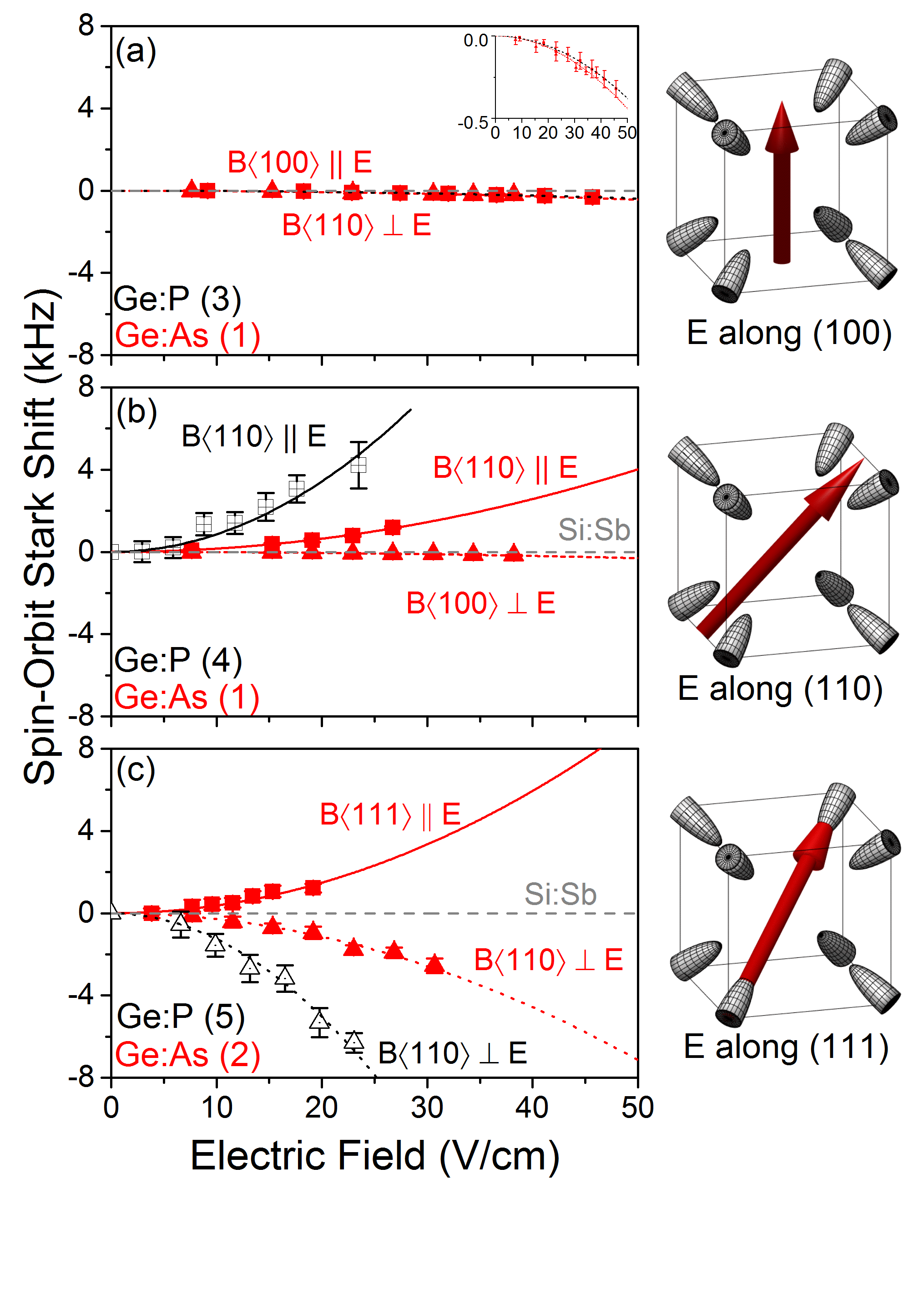}

\caption{\label{fig:fig3exp} Spin-orbit Stark shift for $^{75}$As (red, solid symbols) and $^{31}$P (black, open symbols) donors in germanium. The sample number is listed in each panel's legend and corresponds to the number listed in Table 1. The square symbols with solid lines denote $\vec{E} \parallel \vec{B}$ and triangular symbols with broken lines denote $\vec{E} \perp \vec{B}$. The cartoons to the right schematically show the electric field (red arrow) relative to the conduction band valleys (grey ellipsoids). In (a) the electric field makes equal angles with all of the conduction band valleys so only the ''single-valley" Stark effect contributes to the shift. The inset shows that although small, the Stark effect is resolved in this orientation. In (b) the electric field is oriented between two valleys and in (c) the electric field is directed along one valley axis. When E is along the valley axis, the valley repopulation effect should be maximized. The lines plotted are least squares fits to the data with the exception of the nearly horizontal dashed gray line which represents the strongest Stark shift measured for donors in silicon (hyperfine shift of Si:Sb, $M_{I} = 5/2$).}
\end{figure}

\begin{table*}[h]
 \centering
 \begin{tabular}{|l|c|c|c|r|r|r|r|}
 \hline
 Donor&\multicolumn{1}{c|}{$\vec{E}$ Orientation} & \multicolumn{1}{c|}{$\perp$ / $\parallel$}& \multicolumn{1}{c|}{$\vec{B}_{0}$ Orientation} & \multicolumn{1}{c|}{ $\eta_{a}$($\mu m^{2}$/$V^{2}$)} & \multicolumn{1}{c|}{$\eta_{a}$ theory ($\mu m^{2}$/$V^{2}$)} & \multicolumn{1}{c|}{$\eta_{g}$($\mu m^{2}$/$V^{2}$)} & \multicolumn{1}{c|}{$\eta_{g}$ theory($\mu m^{2}$/$V^{2}$)}\\
 \hline
 \parbox[t]{2mm}{\multirow{6}{*}{$^{75}$As}} 
 & $[$001$]$ &$\perp$& $[110]$ & $(-1.3 \pm 0.1) \times 10^{-1}$ & $-1.2 \times 10^{-1}$ & $  (-1.8 \pm 0.1) \times 10^{-3}$ & ...  \\ % inserting body of the table
 & $[$001$]$ &$\parallel$& $[$001$]$ & $(-8.2 \pm 0.9) \times 10^{-2}$ & $-1.2\times 10^{-1}$ &$ (-1.6 \pm 0.1) \times 10^{-3}$ & ... \\
 & $[$110$]$ &$\perp$& $[001]$ & $(-7.8 \pm 1.5) \times 10^{-2}$& $-9.6\times 10^{-2}$ &$ (-1.3 \pm 0.1) \times 10^{-3}$ &$-1.7 \times 10^{-2}$  \\
 & $[$110$]$ &$\parallel$& $[$110$]$ & $. . .$ & $-9.6\times 10^{-2}$ &$ (1.7 \pm 0.1) \times 10^{-2}$ & $1.7 \times 10^{-2}$  \\
 & $[\overline{1}11]$ &$\perp$&  $[01\overline{1}]$ & $ . . .  $ & $-1.2 \times 10^{-1}$ &$ (-3.0 \pm 0.2) \times 10^{-2}$ &  $-2.0 \times 10^{-2}$ \\
 & $[\overline{1}11]$ &$\parallel$& $[\overline{1}11]$ & $ . . . $ & $-1.2 \times 10^{-1}$ & $ (3.9 \pm 0.4) \times 10^{-2}$ &  $ 4.0 \times 10^{-2}$  \\ [1ex] % [1ex] adds vertical space
\hline

\parbox[t]{2mm}{\multirow{3}{*}{{$^{31}$P}}} 
 & $[$100$]$ &$\parallel$&$[100]$ & $ (-2.2 \pm 0.1) \times 10^{-1}$ & $-2.4\times 10^{-1}$ & $ (-1.3 \pm 0.3) \times 10^{-3}$ & $-4.8\times 10^{-3}$*\\
 & $[$110$]$ &$\parallel$&$[110]$ & $. . .$ & $-2.1\times 10^{-1}$ &$ (9.0 \pm 1.1) \times 10^{-2}$ & $1.0 \times 10^{-1}$\\
 & $[$111$]$ &$\perp$& $[01\overline{1}]$ & $. . .$ & $-2.7\times 10^{-1}$ &$ (-1.3 \pm 0.1) \times 10^{-1}$ & $-9.5 \times 10^{-2}$\\
 \hline
 \end{tabular}
 \caption{\label{table:fittingparams} % is used to refer this table in the text
 Hyperfine ($\eta_{a}$) and spin-orbit ($\eta_{g}$) Stark parameters for $^{31}$P and $^{75}$As donors extracted from the data in Figs.\ref{fig:fig2exp} and \ref{fig:fig3exp}. The theoretical value marked with (*) is taken from \cite{rahman2009} and all other theoretical values are courtesy of Pica \textit{et al.} \cite{picainprep}. These theories match nicely with the experimental results. Note that the Stark parameters are highly anisotropic, changing sign and amplitude by more than an order of magnitude depending on the electric and magnetic field orientations. The Stark parameters are largest for the $^{31}$P donors which are shallower than $^{75}$As donors.}
 \end{table*}

The highly anisotropic spin-orbit Stark shift is shown in Fig.~\ref{fig:fig3exp} for various electric and magnetic field orientations. When the electric field is oriented in the (100) crystallographic directions (Fig.~\ref{fig:fig3exp}(a)), the shift is solely due to the single-valley Stark effect and is small. When the electric field is oriented along the (110) or (111) directions (Fig.~\ref{fig:fig3exp} (b) and (c)), valley repopulation also contributes to the Stark shift, and we see that the shift is up to two orders of magnitude larger. We thus conclude that valley repopulation is the dominant mechanism contributing to the spin-orbit Stark shift. To emphasize the anisotropy in the Stark shift, all three panels are plotted on the same scale. We note that while this makes it difficult to resolve the Stark shift in Fig. ~\ref{fig:fig3exp}(a), the same data are plotted in Fig.~\ref{fig:fig2exp}. The data were least-squares fit with Eq.~3 (neglecting the $\eta_{A}$ term, which was averaged out), and we extract the spin-orbit Stark parameters as recorded in Table~2. The error reported in the table represents the fitting error.

Random strain can also lead to errors in measuring the hyperfine and spin-orbit Stark parameters since it is equivalent to internal electric fields ($\vec{E}_{int}$). When ($\vec{E}_{int}$) is superimposed with our externally applied electric field ($\vec{E}_{ext}$), it can lead to a large linear Stark effect since the Stark shift is then proportional to $(\vec{E}_{int} + \vec{E}_{ext})^{2}$. This linear term was cancelled by applying bipolar electric field pulses as previously discussed\cite{bradbury2006}.

To compare these shifts with what was reported for donors in silicon, we plot the largest Stark shift measured for donor electron spins in silicon (the $M_{I} = 5/2$ transition for $^{121}$Sb donors) \cite{pica2014} in Fig.~\ref{fig:fig3exp}. This shift is colored gray and is so small that it appears flat. At a field of 50 V/cm, the shift for Si:Sb is only $\sim$ -3 Hz, compared to over 9 kHz for Ge:As with a $(111)$ oriented $\vec{E} \parallel \vec{B}_{0}$. From these data, it is clear that in terms of Stark sensitivity, germanium far outperforms silicon.

Of course, high sensitivity does not necessarily translate into large tunability. For the donors in a large ensemble to be gate addressable, one would like to be able to apply large enough electric fields to reliably tune the donor electron spin by more than the ensemble linewidth. In our recent work \cite{SigillitoGe2015}, we have found that the ensemble linewidth of donor electron spins in highly enriched germanium can be as narrow as 1.1 MHz (0.05 mT). With the electric fields applied in this work, we were able to demonstrate a Stark shift of only 7 kHz (Fig.~\ref{fig:fig3exp} (c)). The largest electric field was limited by the high densities of $^{31}$P and $^{75}$As donors in our samples, which can undergo avalanche impact ionization at higher fields given the large separations between the parallel plates \cite{steele1959}. Much larger electric fields will be permitted in nano-scale gated devices or in lightly doped macroscopic crystals. In the supplementary information, we demonstrate that fields as large as 480 V/cm can be applied to a 0.5 mm thick crystal with $\sim 10^{12}$ $^{31}$P/cm$^{3}$ without signs of donor ionization. The resulting Stark shift is 28 kHz and is relatively small because electric fields could only be applied along a (100) crystallographic axis. However, a similar non-ionizing electric field of 480 V/cm applied along the (111) direction, would produce a Stark shift of 4.2 MHz (For $\vec{B}_{0}$ parallel to $\vec{E}$), exceeding the ensemble linewidth (of $0.01\% \ ^{73}$Ge) by a factor of four \cite{picainprep}.

Because spins can be tuned by more than the ensemble linewidth, donors in germanium are compatible with Stark addressable spin manipulation schemes. Stark modulation was demonstrated for individual spins in silicon where the "instantaneous" spin linewidth is narrow\cite{lauchte2015}. In this work, a field of 8000 V/cm was applied to achieve a shift of 350 kHz. The large shift was made possible by a very large linear Stark effect, presumably due to strain in their nano-scale gated devices. Addressability was achieved for ensembles of Sb nuclear spins, which have very narrow linewidths \cite{wolfowicz2014nuc}. A field of 900 V/cm was used in this work to produce a shift of 8 kHz. These large electric fields are not necessary in germanium.

\section{Conclusion}

In summary, we have investigated the Stark tunability of $^{31}$P and $^{75}$As donor qubits in germanium--a largely unstudied quantum system that offers some major advantages over silicon. Our results show that the spin-orbit and hyperfine components of the Stark shift are four orders and one order of magnitude larger, respectively, when compared with silicon. We find a lower bound for ionizing fields in our enriched samples of 480 V/cm, which gives a lower limit on the Stark tunability of Ge donor qubits of 4.2 MHz, four times the ensemble linewidth (1.1 MHz \cite{SigillitoGe2015}). This means that even large ensembles of donor qubits in germanium can be reliably gated using electric fields. When these encouraging results are combined with the long coherence times we have already reported \cite{SigillitoGe2015} and germanium's compatibility with industrial semiconductor processing \cite{shang2003, yu2011, kim2014, scappucci2011}, germanium appears to be the natural host material for the next generation of donor-based quantum bits.

\begin{acknowledgments}
We thank Giuseppe Pica and Brendon Lovett for fruitful discussions and access to their unpublished calculations. We would also like to thank Cheuk Chi Lo for the use of his natural germanium samples. Work at Princeton was supported by the NSF through the Princeton MRSEC (DMR-01420541), and by the ARO (W911NF-13-1-0179). The work at Keio was supported by KAKENHI (S) Grant No. 26220602, and JSPS Core-to-Core, and Spintronics Research Network of Japan.

\end{acknowledgments}
\providecommand{\noopsort}[1]{}\providecommand{\singleletter}[1]{#1}%

\end{document}